\documentclass{SciPost}
\pdfoutput=1

\usepackage{gensymb}
\usepackage{braket,slashed,bm}
\usepackage{booktabs}
\usepackage{array,multirow}
\usepackage{amsmath}
\usepackage[normalem]{ulem}
\usepackage[T1]{fontenc}
\usepackage{mathrsfs}
\usepackage{tikz}
\usepackage{stmaryrd}
\usepackage{subcaption}
\usetikzlibrary{arrows,decorations.pathmorphing,backgrounds,positioning,fit,petri,automata,shadows,calendar,mindmap,
decorations.markings,calc}
\usepackage{lscape}
\usepackage{hyperref}
\usepackage{graphicx,wrapfig}
\usepackage{floatrow}
\newfloatcommand{capbtabbox}{table}[][\FBwidth]

\def\bea{\begin{eqnarray}}
\def\eea{\end{eqnarray}}

\begin{document}

\begin{center}{\Large \textbf{
Statistics of predictions with missing higher order corrections
}}\end{center}

\begin{center}
Laure Berthier\textsuperscript{1*}, 
Jeppe Tr{\o}st Nielsen\textsuperscript{1}
\end{center}

\begin{center}
{\bf 1} Niels Bohr International Academy, Blegdamsvej 17, Copenhagen, Denmark
\\
* berthier@nbi.ku.dk
\end{center}

\begin{center}
\today
\end{center}


\section*{Abstract}
{\bf 
Effective operators have been used extensively to understand small deviations from the Standard Model in the search for new physics. So far there has been no general method to fit for small parameters when higher order corrections in these parameters are present but unknown. We present a new technique that solves this problem, allowing for an exact p-value calculation under the assumption that the leading higher order theoretical contributions can be treated as gaussian distributed random variables and all remaining higher orders can be neglected. The method we propose is general, and may be used in the analysis of any perturbative theoretical prediction, ie.~truncated power series. We illustrate this new method by performing a fit of the Standard Model Effective Field Theory parameters, which include eg.~anomalous gauge and four-fermion couplings. 
}

\vspace{10pt}
\noindent\rule{\textwidth}{1pt}
\tableofcontents\thispagestyle{fancy}
\noindent\rule{\textwidth}{1pt}
\vspace{10pt}

\section{Introduction}\label{sec:intro}
The discovery of the Higgs boson at the LHC \cite{Aad:2012tfa,Chatrchyan:2012xdj}, and the subsequent lack of any new resonances, have put an emphasis on precision studies of collider data. A general approach to this is to parametrise possible deviations from the Standard Model (SM) by effective operators. Historically, only oblique corrections were considered in the study of the electroweak data~\cite{Peskin:1990zt,Peskin:1991sw,Kennedy:1991sn}. More recently, the experiments at the LHC have considered contributions from effective operators to vertices when looking at corrections to the Higgs data~\cite{Aad:2015pla,Aad:2015tna,Khachatryan:2016tnr} and in search of anomalous gauge couplings~\cite{Aad:2016sau,Aad:2016ett,Aad:2016wpd,Khachatryan:2015kea,Khachatryan:2015sga,Khachatryan:2016yro}. Of such approaches the most general is to include all higher dimension operators allowed by gauge invariance as it is done in the SM effective field theory (SMEFT). Comprehensive studies of electroweak precision data~\cite{Grinstein:1991cd,Han:2004az,Berthier:2015gja,Berthier:2015oma,Berthier:2016tkq}, LHC data~\cite{Dumont:2013wma,Ciuchini:2014dea,Falkowski:2015fla,Falkowski:2015jaa} and top quark data~\cite{Buckley:2015nca,Buckley:2015lku} have now put constraints on a number of parameters in the SMEFT. An aspect of these calculations which so far has received little attention is the consistent statistical treatment of missing higher order corrections from the parameters included in the fit. Usually the predictions in the effective field theory are calculated only to first order in the Wilson coefficients, and so contributions from eg.~operators squared are left uncertain in the theoretical computation. In previous works, these uncertainties from missing higher orders have either been included as a simple constant error or not at all. These have the common problem that they produce a $\Delta \chi^2$ statistic which is not in general chi-squared distributed. This means that the computed confidence levels cannot be trusted. Concerns about the validity of the statistical methods used at the LHC for anomalous gauge coupling fits have also been raised in \cite{Gregersen:2015uea}, although they address a somewhat different problem than we do.

This problem is generic in the sense that any truncated power series has corrections which depend on the expansion parameter. To consistently constrain this expansion parameter, we must include an error, which in turn depends on the parameter we are trying to determine, ie.~the error itself becomes inherently unknown. That is the problem we address in the present work. Our proposed solution includes this theoretical uncertainty in such a way that the constructed $\Delta\chi^2$ recovers the usual chi-squared distribution. As a special case, if we are only concerned with the statistical evidence for a non-zero value of a parameter ie.~the p-value at the origin of the parameter space, this method suggests one should not include any extra theoretical uncertainty. This in particular means that including a fixed theoretical error unnecessarily worsens the determination of the existence of new physics. To construct confidence regions, the correct inclusion of the theoretical uncertainty is straightforward, however more involved than the standard treatment  and can lead to different conclusions than the inclusion of a constant uncertainty.

Section~\ref{sec:fit} defines the problem more precisely and presents its solution, and Sec.~\ref{sec:application} shows an application of this method to the analysis of the SMEFT. Section~\ref{sec:con} concludes and summarises our results.
\section{Statistics of missing higher orders} \label{sec:fit}
In the present section we formulate the problem of fitting small parameters with missing higher order contributions from the same parameters and our solution to this problem. The solution given here is for observables calculated to first order, however it can in principle be extended to any order. 

We consider $N$ observables $O_i$, which can be treated perturbatively in $M$ parameters $\alpha_j$, which we wish to determine. Given only the first order, we have in general
\begin{align}
O_i(\alpha) = O^{(0)}_i + \sum_j O^{(1)}_{ij}\alpha_j  + \mathcal{O}(\alpha^2)\, ,
\end{align}
where the theoretical uncertainty is of the order $\alpha^2$. The problem is how to consistently include the theoretical uncertainty in the statistical treatment of the data. Consider the missing second order terms to the prediction $O_i(\alpha)$, which is $\sum_{jk}O^{(2)}_{ijk}\alpha_j\alpha_k$. Since the coefficients $O^{(2)}_{ijk}$ are unknown, we pretend they can be treated as random variables. While they obviously are not random -- they are after all given if one were to do the calculation -- we may, for the purposes of error estimation, treat them as such. We therefore ascribe a variance $\sigma_{TH,ijk}$ to each higher order coefficient $O^{(2)}_{ijk}$, which is a rough estimate of that coefficient.\footnote{Estimating $\sigma_{TH}$ is a problem on its own, which is quite independent from the problem at hand. Determining the best estimate depends entirely on the theory under consideration. For a discussion of this in the context of QCD, chiral perturbation theory and Higgs effective field theory see eg.~\cite{Bagnaschi:2014wea,Cacciari:2011ze,Isgro:2015cfa,Furnstahl:2015rha,Fichet:2015xla}. We will in the next section discuss this term in the context of the SMEFT.} 
Suppose first we do know the true theoretical uncertainty $\sum_{jk}\sigma_{TH,ijk}\alpha_{true,j}\alpha_{true,k}$, including the dependence on the true value of $\alpha$, which we write as $\alpha_{true}$. 
Further assuming the error is gaussian, we write down a normal $\Delta\chi^2$ statistic, assuming \emph{without loss of generality} no covariant errors,
\begin{align}\label{eq:statTH}
\chi^2_{TH}(\alpha) &=  \sum_i \frac{[ \hat O_i - O_i(\alpha)]^2}{\sigma_i^2 + \sum_{j,k}\sigma_{TH,ijk}^2\alpha_{true,j}^2\alpha_{true,k}^2} \\
\Delta\chi^2_{TH}(\alpha) &= \sum_i \frac{[ \hat O_i - O_i(\alpha)]^2}{\sigma_i^2 + \sum_{j,k} \sigma_{TH,ijk}^2\alpha_{true,j}^2\alpha_{true,k}^2} 
- \sum_i \frac{[ \hat O_i - O_i\left(\hat\alpha\right)]^2}{\sigma_i^2 + \sum_{j,k} \sigma_{TH,ijk}^2\alpha_{true,j}^2\alpha_{true,k}^2} \, ,
\label{eq:statdTH}\end{align}
where $\hat O_i$ is the data, $\sigma_i$ is the experimental uncertainty and $\sum_{jk}\sigma_{TH,ijk}\alpha_{true,j}\alpha_{true,k}$ is the true theoretical uncertainty.
We denote by $\hat{\alpha}$ the value of $\alpha$ that minimizes $\chi^2_{TH}(\alpha)$.
Following standard procedures, this statistic evaluated at $\alpha_{true}$ is a random variable with a chi-squared distribution and we construct confidence regions as contours of constant $\Delta\chi^2_{TH}(\alpha)$. In a frequentist setting we are only ever concerned with the value of the $\Delta\chi^2_{TH}$ at the true $\alpha$ -- the confidence interval either does or does not contain the true parameters, and that is determined solely by the value of the statistic there. 

Obviously, the above construction cannot be done in practice. However, we may construct a statistic, which has the same value at the true parameter, but in general is different everywhere else.
We define the $\chi^2$, separating its $\alpha$ dependence in the numerator $\alpha_N$ and denominator $\alpha_D$
\bea
\chi^2(\alpha_N, \alpha_D) = \sum_i \frac{[ \hat O_i - O_i(\alpha_N)]^2}{\sigma_i^2 + \sum_{j,k} \sigma_{TH,ijk}^2\alpha_{D,j}^2\alpha_{D,k}^2}. 
\eea
By not having just one minimal value of the $\chi^2$, but one minimum \emph{for every choice of $\alpha_D$}, we can emulate the behavior of $\Delta\chi^2_{TH}$.\footnote{This may of course be computationally demanding  for very involved likelihood calculations. In particular it requires the inversion of a new covariance matrix for every evaluation. For the datasets we consider here, this is of no practical importance. For larger datasets, one may have to settle for an approximate scheme.}
Denoting by $\hat \alpha(\alpha)$ the value of $\alpha_N$ that minimises the $\chi^2$ for the particular choice $\alpha_D=\alpha$ in the sum $\sum_{j,k}\sigma_{TH,ijk}^2 \alpha_{D,j}^2\alpha_{D,k}^2 $, we construct our proposed statistic
\begin{align}\label{eq:stat}
\Delta\chi^2(\alpha) =&\, \chi^2 (\alpha, \alpha) - \chi^2 (\hat{\alpha}(\alpha),\alpha) \nonumber \\
=& \sum_i \frac{[ \hat O_i - O_i(\alpha)]^2}{\sigma_i^2 + \sum_{j,k} \sigma_{TH,ijk}^2\alpha_j^2\alpha_k^2} 
- \sum_i \frac{[ \hat O_i - O_i\left(\hat\alpha(\alpha)\right)]^2}{\sigma_i^2 + \sum_{j,k} \sigma_{TH,ijk}^2\alpha_j^2\alpha_k^2} 
 \,, 
\end{align}
It is immediate that inserting $\alpha_{true}$, the minimizing $\hat\alpha(\alpha_{true})$ will be the $\hat\alpha$ of Eq.~\ref{eq:statdTH}.
Therefore this new statistic will satisfy $\Delta\chi^2(\alpha_{true})=\Delta\chi^2_{TH}(\alpha_{true})$, and as a direct consequence the proposed statistic $\Delta\chi^2(\alpha_{true})$ follows a chi-squared distribution. That in turn means that any confidence interval constructed from the new $\Delta\chi^2$ will contain the true value of the parameters if and only if it is contained in the imagined confidence interval constructed from $\Delta\chi^2_{TH}$ -- which is constructed in a completely standard fashion. Therefore, the confidence levels derived from this new $\Delta\chi^2$ are strictly in accordance with the standard construction, only the resulting confidence regions are different. The coverage properties of the confidence regions are necessarily identical.
Equation~\eqref{eq:stat} defining our statistic and the fact that it is chi-squared distributed are our main results.

Extending the previous procedure to non-linear models, eg. only missing third or higher orders, the distribution of the $\Delta\chi^2$ is expected to follow a chi-squared distribution only asymptotically. This statement, as dictated by Wilks' theorem, holds for any non-linear model. Our extension with theoretical errors does not change this fact. However, for the linear model with an extra uncertainty we have considered so far, this is not relevant. In this case, the theoretical error, no matter its size, does not make the assumed model non-linear.

We can think of this statistic in terms of hypothesis testing. We wish to know at every point $\alpha$ in parameter space, what is the probability to exceed the computed $\Delta\chi^2$ given $\alpha$ are the true parameters. This demands that the theoretical uncertainty \emph{even when computing the minimum} is given in terms of the parameters $\alpha$, and \emph{not} $\hat\alpha(\alpha)$. If we instead simply minimise a $\chi^2(\alpha)$ with the uncertainties free, we immediately lose the strict chi-squared distribution at the true value. 

For the simple example of a linear model with covariant errors, we can explicitly write down the minimising parameters. Call the covariance matrix of the observables $V$, which includes the theoretical uncertainty. In vector notation the minimising $\hat\alpha(\alpha)$ becomes
\bea
\hat \alpha(\alpha) = ( O^{(1)T} V(\alpha)^{-1} O^{(1)})^{-1}O^{(1)T} V(\alpha)^{-1} (\hat O - O^{(0)}) \label{MLE} \,,
\eea
where $\hat O, O^{(0)}, O^{(1)}$ are understood as two vectors and a matrix respectively. Notice the dependence on $\alpha$ here is only through the covariance matrix. The $\Delta\chi^2$ statistic is then given by
\bea
\Delta\chi^2(\alpha) = (\alpha - \hat\alpha(\alpha)) O^{(1)T}V(\alpha)^{-1}O^{(1)} (\alpha - \hat\alpha(\alpha)) \label{Delta_Chi} \,.
\eea

It should be noted that this method \emph{cannot} include nuisance parameters with missing higher orders, eg.~by profiling. This traces back to the fact that the original statistic in Eq.~\eqref{eq:statTH} would be minimised over a subset of the $\alpha$'s, while keeping the theoretical error fixed. We cannot do this with Eq.~\eqref{eq:stat} since minimising over this subset also changes the $\alpha$ determining the theoretical uncertainty. As we will usually be interested in all parameters of this type, we do not see this as a big drawback. Only in illustrating the constraints does this really come into play. Of course there is no problem in simply putting some parameters to zero or some other fixed value, as this is effectively ignoring the dependence completely or in other words, working in a different theory. This is indeed what we do to illustrate the effect of the procedure in the next section.

We have also verified by a Monte Carlo study that the $\Delta\chi^2$ of Eq.~\eqref{eq:stat} is chi-squared distributed. An important point to keep in mind when doing this study is to treat the theoretical error as a source of noise. As such all the coefficients $O^{(2)}_{ijk}$ need to be re-drawn from their respective gaussian distributions for every simulation. The simulated experimental data for the $n$th simulation is then
\begin{equation}
\hat O_i^{[n]} = O^{(0)}_i + \sum_j O^{(1)}_{ij}\alpha_{true,j} + \sum_{j,k} x^{[n]}_{ijk}\alpha_{true,j}\alpha_{true,k} + y^{[n]}_i \, ,
\end{equation}
where $x^{[n]}_{ijk}$ is a gaussian random variable with standard deviation $\sigma_{TH,ijk}$ and $y^{[n]}_i$ is the experimental noise with standard deviation $\sigma_i$.
\section{Application to the Standard Model Effective Field Theory}  \label{sec:application}
In this section, we apply our new fitting method to the SMEFT. We then compare the results obtained with the results of other procedures. 

We consider the linear SMEFT in which the Higgs scalar is embedded in a doublet that belongs to the fundamental representation of $SU(2)_L \times U(1)_Y$. The SMEFT Lagrangian is given by the sum of the SM Lagrangian and a series of higher ($>\hspace{-1.5mm}4$) dimension operators suppressed by the appropriate power of the cutoff scale, 
\bea
\mathcal{L}_{SMEFT} = \mathcal{L}_{SM} + \sum \limits_{k=5}^{\infty} \mathcal{L}^{(k)},
\eea 
where 
\bea
\mathcal{L}^{(k)} = \sum \limits_{i} \frac{C^{(k)}_i}{\Lambda^{k-4}}Q^{(k)}_i,
\eea
where $Q^{(k)}_i$ are the operators of dimension $k$ with the associated Wilson coefficient $C^{(k)}_i$ and $\Lambda$ is the cutoff-scale. In what follows, we consider only dimension-6 operators. The Lagrangian is then simply $ \mathcal{L}_{SM} + \mathcal{L}^{(6)}$, and we use the Warsaw basis for the dimension-6 operators \cite{Grzadkowski:2010es}. 


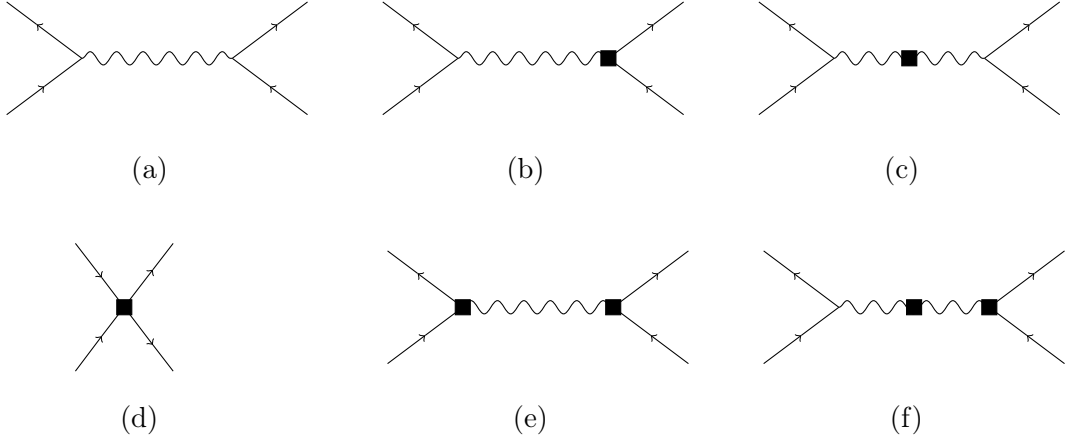
\begin{figure}[tb]

\begin{tikzpicture}
\hspace*{-5cm}

\draw[decorate,decoration=snake] (-0.5,0) -- (1.5,0) ;
\draw [->](1.5,0) -- (2.1,0.45);
\draw (2.1,0.45) -- (2.5,0.75);
\draw (1.5,0) -- (2,-0.375);
\draw [<-](2,-0.375) -- (2.5,-0.75);

\draw [->](-0.5,0) -- (-1.1,0.45);
\draw (-1.1,0.45) -- (-1.5,0.75);
\draw (-0.5,0) -- (-1,-0.375);
\draw [<-](-1,-0.375) -- (-1.5,-0.75);

\draw (0.4,-1.5) node [align=center] {(a)};

\hspace{5cm}
\draw[decorate,decoration=snake] (-0.5,0) -- (1.5,0) ;
\filldraw (1.4,-0.1) rectangle (1.6,0.1);
\draw [->](1.5,0) -- (2.1,0.45);
\draw (2.1,0.45) -- (2.5,0.75);
\draw (1.5,0) -- (2,-0.375);
\draw [<-](2,-0.375) -- (2.5,-0.75);

\draw [->](-0.5,0) -- (-1.1,0.45);
\draw (-1.1,0.45) -- (-1.5,0.75);
\draw (-0.5,0) -- (-1,-0.375);
\draw [<-](-1,-0.375) -- (-1.5,-0.75);

\draw (0.4,-1.5) node [align=center] {(b)};

\hspace{5cm}

\draw[decorate,decoration=snake] (-0.5,0) -- (1.5,0) ;
\filldraw (0.4,-0.1) rectangle (0.6,0.1);
\draw [->](1.5,0) -- (2.1,0.45);
\draw (2.1,0.45) -- (2.5,0.75);
\draw (1.5,0) -- (2,-0.375);
\draw [<-](2,-0.375) -- (2.5,-0.75);

\draw [->](-0.5,0) -- (-1.1,0.45);
\draw (-1.1,0.45) -- (-1.5,0.75);
\draw (-0.5,0) -- (-1,-0.375);
\draw [<-](-1,-0.375) -- (-1.5,-0.75);
\draw (0.4,-1.5) node [align=center] {(c)};

\end{tikzpicture}
\vspace{0.6cm}

\begin{tikzpicture}
\hspace*{-5cm}
\draw[->] (-0.65,-0.85) -- (-0.3,-0.385);
\draw [->](-0.3,-0.385) -- (0.38,0.5) ;
\draw (0.38, 0.5) --(0.65,0.85);
\draw [->](-0.65,0.85) -- (-0.3,0.395);
\draw [->](-0.3, 0.395) -- (0.38,-0.5) ;
\draw  (0.38,-0.5)--(0.65,-0.85);
\filldraw (-0.1,-0.1) rectangle (0.1,0.1);

\draw (0.2,-1.5) node [align=center] {(d)};

\hspace{5cm}
\draw[decorate,decoration=snake] (-0.5,0) -- (1.5,0) ;
\filldraw (-0.4,-0.1) rectangle (-0.6,0.1);
\filldraw (1.4,-0.1) rectangle (1.6,0.1);
\draw [->](1.5,0) -- (2.1,0.45);
\draw (2.1,0.45) -- (2.5,0.75);
\draw (1.5,0) -- (2,-0.375);
\draw [<-](2,-0.375) -- (2.5,-0.75);

\draw [->](-0.5,0) -- (-1.1,0.45);
\draw (-1.1,0.45) -- (-1.5,0.75);
\draw (-0.5,0) -- (-1,-0.375);
\draw [<-](-1,-0.375) -- (-1.5,-0.75);

\draw (0.4,-1.5) node [align=center] {(e)};

\hspace{5cm}
\draw[decorate,decoration=snake] (-0.5,0) -- (1.5,0) ;
\filldraw (0.4,-0.1) rectangle (0.6,0.1);
\filldraw (1.4,-0.1) rectangle (1.6,0.1);
\draw [->](1.5,0) -- (2.1,0.45);
\draw (2.1,0.45) -- (2.5,0.75);
\draw (1.5,0) -- (2,-0.375);
\draw [<-](2,-0.375) -- (2.5,-0.75);

\draw [->](-0.5,0) -- (-1.1,0.45);
\draw (-1.1,0.45) -- (-1.5,0.75);
\draw (-0.5,0) -- (-1,-0.375);
\draw [<-](-1,-0.375) -- (-1.5,-0.75);

\draw (0.4,-1.5) node [align=center] {(f)};


\end{tikzpicture}

\caption{Feynman diagrams contributing to $e^+ e^- \rightarrow f \bar{f}$ through a $Z$ in the SMEFT. The box indicates the insertion of Wilson coefficients of dimension-6 operators. These Wilson coefficients appear, for examples, in the redefinition of the Z couplings to fermions (b), in the redefinition of the Z gauge boson width (c) or generates four-fermion interactions (d). The cross section in the SMEFT is proportional to the square of the sum of these diagrams. Interferences between (a) and (b), (c), (d) generate the first order correction to the cross section. The second order terms are given by the interferences between pairs of the diagrams (b), (c), (d) and between (a) and (e) or (f). There are also second order corrections to the vertices and propagators meaning that even interference between (a) and (b), (c) has second order terms. It should be understood that the insertion of Wilson coefficients can happen in the other vertex in (b) and (f) and these diagrams have to be taken into account when calculating the amplitude. To calculate the full cross section of the process $e^+ e^- \rightarrow f \bar{f}$,  we add to the diagrams already drawn here the s-channel diagrams through a $\gamma$-exchange in the SMEFT. }
\label{Diagrams}

\end{figure}


We consider the 176 electroweak observables described in \cite{Berthier:2015oma,Berthier:2016tkq}. This set of observables include the LEP1 pseudo-observables \cite{ALEPH:2005ab}, LEP2 data \cite{Schael:2013ita}, data from older colliders (TRISTAN, PEP, PETRA)~\cite{Inoue:2000hc,Velissaris:1994rv,Miura:1997mq,Hegner:1989rd,Derrick:1985gs,Sagawa:1996yf,Abe:1993xr,Gan:1985st}, low energy precision measurements \cite{Zeller:2001hh,Allaby:1987vr,Vilain:1994qy,Ahrens:1990fp,McFarland:1997wx,Anthony:2005pm,Prescott:1979dh,Dorenbosch:1988is,Vetter:1995vf,Wood:1997zq,Ito:2003mr} as well as the $e^+e^- \rightarrow W^+W^- \rightarrow 4f$ total and differential cross sections measured at LEP2 \cite{Achard:2004zw,Abbiendi:2007rs,Heister:2004wr,Schael:2013ita}. In the SMEFT, these observables receive corrections from dimension-6 operators. In total, 20 dimension-6 operators contribute to the redefinition of the observables we consider. We adopt similar notations as in Sec.~\ref{sec:fit}. We denote by $\hat{O}$ the vector of the measured values of the observables and by $O$ the vector of the SMEFT predictions of the observables at first order in the dimension-6 operators. The value of the observable $O_i$ is
\bea
O_i = O^{(0)}_i + \sum \limits_{j} O^{(1)}_{ij} \frac{C_j}{\Lambda^2}\label{eq:Oi}
\eea
where the $C$ is the vector of $C^{(6)}$ contributing to the observables and is given by
\bea
C =  \{ C_{He}, C_{Hu}, C_{Hd}, C_{H l}^{(1)}, C_{Hl}^{(3)}, C_{Hq}^{(1)}, C_{Hq}^{(3)}, C_{HWB}, C_{HD},\nonumber \\ C_{ll}, C_{ee}, C_{eu}, C_{ed}, C_{le}, C_{lu}, C_{ld}, C_{lq}^{(1)}, C_{lq}^{(3)}, C_{qe}, C_W\}. 
\eea
Note that $O^{(0)}$ is a 176 dimensional vector and $O^{(1)}$ a $176\times 20$ matrix. The second order correction $O_{ijk}^{(2)} C_j C_k / \Lambda^4$ to the SMEFT value of the observable $O_i$ is here unknown. This is exactly a situation in which the method developed in Sec. \ref{sec:fit} applies.

We assume that $O_{ijk}^{(2)}$ can be described as a gaussian random variable with standard deviation 
\bea
\sigma_{TH,ijk}  = O^{(1)}_{ij} O^{(1)}_{ik}/ O^{(0)}_{i}.
\eea
This rough approximation for $\sigma_{TH,ijk}$ can be understood as follows. The numbers $O_{ijk}^{(2)}$ result for instance from the interference between two Feynman diagrams, each having dimension-6 operators inserted. Take, as an example, the scattering $e^+ e^- \rightarrow f \bar{f}$ through a $Z$. The Feynman diagrams contributing to this process in the SMEFT are given in Fig.~\ref{Diagrams} where interference between diagrams generate first and second order corrections to its associated cross section. Denote by $\mathcal{A}_i$ the amplitude in the SMEFT of this process which we write as
\bea
\mathcal{A}_i=\mathcal{A}_{i,SM}\left[1 + \sum_j a_{ij} \frac{C_j}{\Lambda^2} +\sum_{j,k}  b_{ijk} \frac{C_j C_k}{\Lambda^4}  \right]
\eea
where $\mathcal{A}_{i,SM}$ is the amplitude in the SM at tree level and $a_{ij}$, $b_{ijk}$ are numbers that can in principle be calculated in the SMEFT. It follows that the cross section $\sigma_i$ is roughly given by 
\bea
\sigma_i = \sigma_{i,SM}\left[1+\sum_j2a_{ij} \frac{C_j}{\Lambda^2} + \sum_{j,k}(a_{ij} a_{ik}+2b_{ijk}) \frac{C_j C_k}{\Lambda^4} \right] \,.
\eea
Comparing this expression to Eq.~\eqref{eq:Oi}, we read off the numbers $O^{(1)}_{ij}=2a_{ij} O^{(0)}_i$ and  $O^{(2)}_{ijk}= (2b_{ijk}+a_{ij} a_{ik})O^{(0)}_i\approx O^{(1)}_{ij} O^{(1)}_{ik}/ O^{(0)}_i$ for $|b_{ijk}|$ of the order $a_{ij}a_{ik}$, ie.~we estimate the contributions from second order to be of the same size as first order squared. When doing the fit, $O^{(0)}_i$ is taken to be the state-of-the-art (including loop corrections) value of the observable $O_i$ in the SM. The covariance matrix $V$ of the observables includes this theoretical error in the following way
\bea
V_{ij} = V_{SM,ij} + V_{exp,ij} + \delta_{ij} \sum \limits_{k,l} \frac{C^2_{k}C^2_{l}}{\Lambda^8} \sigma_{TH,ikl}^2
\eea
where $V_{SM,ij}$ and $V_{exp,ij}$ are respectively the theoretical SM and experimental covariance matrices and $\sigma_{TH,ikl} = O^{(1)}_{ik}O^{(1)}_{il}/ O^{(0)}_{i}$. Note that we do not include theoretical correlations between observables nor do we discriminate between observables of different nature. A more thorough analysis must consider these more carefully. However this is beyond the scope of the current work. In principle we also have corrections from interference between SM loop corrections and the first order in the SMEFT expansion. These contributions are of the order $(g^2/16\pi^2)O^{(1)}_{ij}C_j\Lambda^{-2}$, and even with a factor $40$ in front, these have little numerical significance compared to the uncertainties we consider.

\begin{figure}[t]
\centering
\includegraphics[width=0.6 \textwidth]{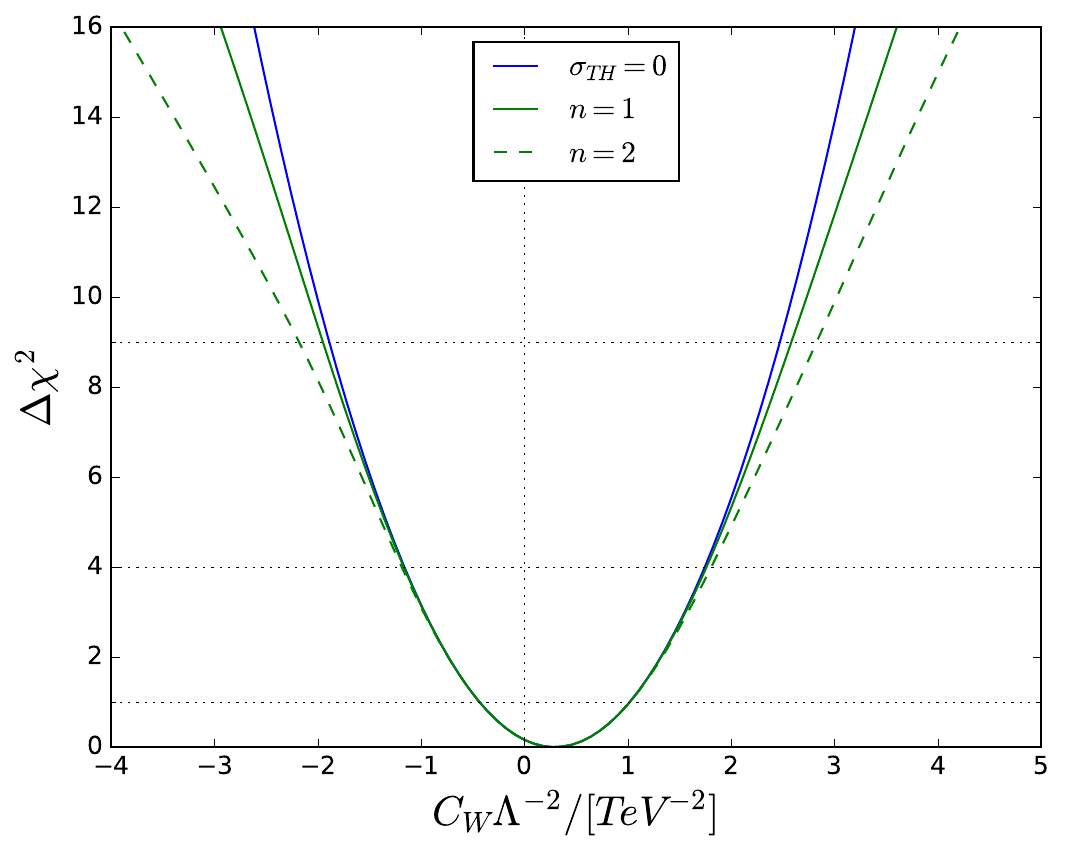}
\caption{$\Delta \chi^2 (C_W \Lambda^{-2})$ for a theoretical uncertainty $n \, \sigma_{TH, ijk}$ for $n=0$ (in blue solid line), $n=1$ (in green solid line) and $n=2$ (in green dashed line). Adding a constant numerical error of the order $10^{-3} O^{(0)}$ barely modifies the $\Delta \chi^2$ -- the change is less than the width of the blue line. In this plot, all other Wilson coefficients are set to 0. \label{n_Dependency}}
\end{figure}

The $\Delta\chi^2$ is easily constructed for all 20 contributing Wilson coefficients using Eq.~\eqref{Delta_Chi}. Its value at the origin, ie.~the SM, is $\Delta\chi^2(0)\approx 27.7$, corresponding to circa $1.5\sigma$, just as quoted in \cite{Berthier:2016tkq}. As stated before, this value is computed explicitly without extra theoretical uncertainty. Since the parameter space is 20 dimensional it is impossible to visualise the constraints on a plot, and we refer to the full $\Delta\chi^2$.

\begin{figure}[t]
\centering
\includegraphics[width=0.6 \textwidth]{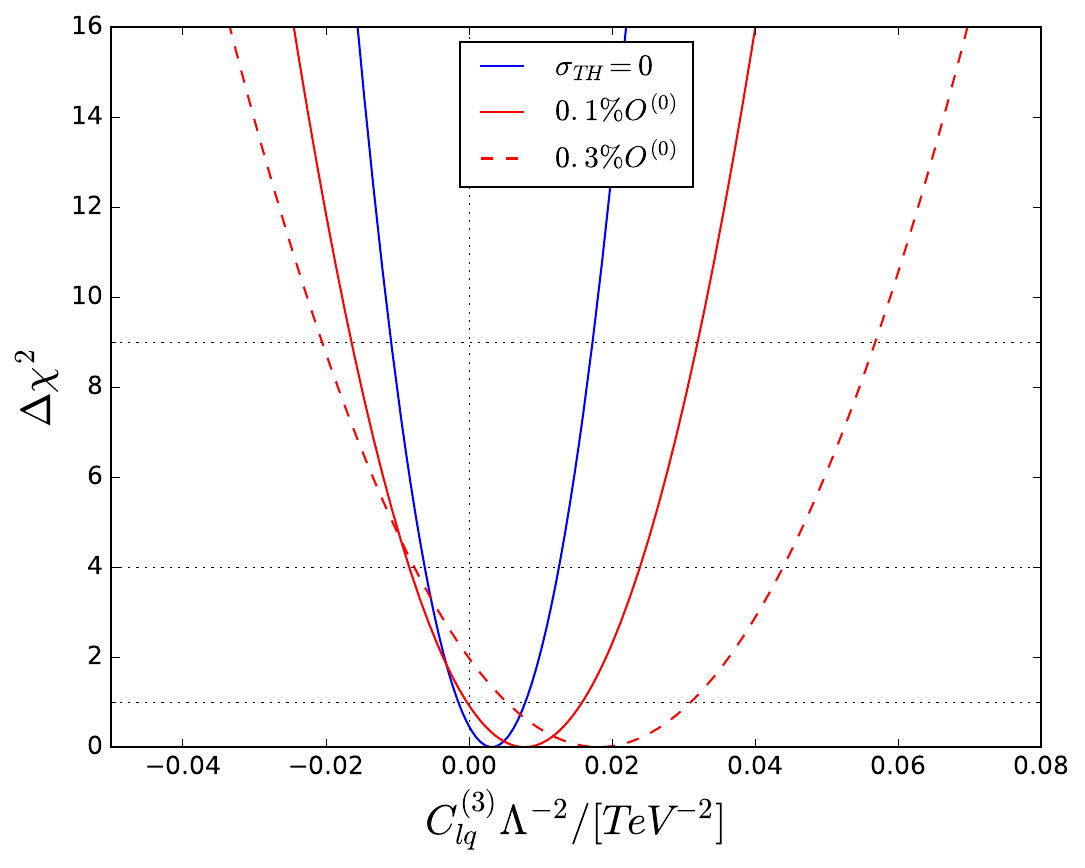}
\caption{$\Delta \chi^2 (C_{lq}^{(3)}\Lambda^{-2})$ when only a constant numerical error of $0\%$ (in blue solid line), $0.1\%$ (in solid red line) and $0.3\%$ (in red dashed line) of $O^{(0)}$ has been added. Adding our estimate of the theoretical uncertainty does not significantly affect the $\Delta \chi^2$ -- the change is less than the width of the blue line. In this fit, other Wilson coefficients are set to 0. 
The changing minimum can be seen as a consequence of arbitrarily changing the degeneracies of the fit from adding a constant error, see the discussion following Eq.~\ref{eq:STdef}.
\label{SMEFT_Dependency}}
\end{figure}

\begin{figure}[t]
\centering
\includegraphics[width=0.6 \textwidth]{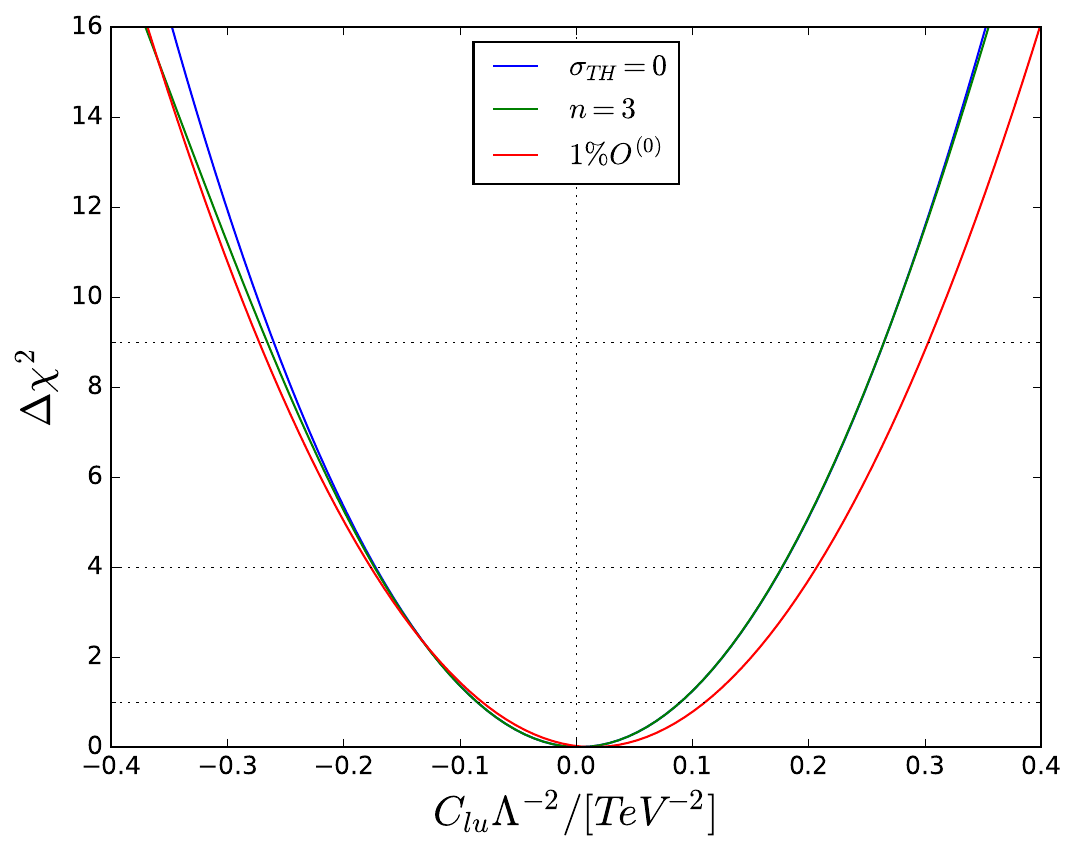}
\caption{$\Delta \chi^2 (C_{lu}\Lambda^{-2})$ when adding no extra error (in blue line), a theoretical error $3 \sigma_{TH, ijk}$ (in green line), and a constant numerical error of $1\% \, O^{(0)}$ (in red line). All other Wilson coefficients are set to 0. \label{Comparison_Method}}
\end{figure}

To visualise the effects of the proposed method, we plot in Figs.~\ref{n_Dependency}, \ref{SMEFT_Dependency} and \ref{Comparison_Method} the constraints on single Wilson coefficients, $C_W, C_{lq}^{(3)}$ and $C_{lu}$ respectively. The fits are made with only a single active Wilson coefficient, ie.~not in the full SMEFT.
These three coefficients are chosen specifically to showcase the different behavior of this method compared to the inclusion of a constant error. We take as an example the constant error to be of the order $10^{-3} O^{(0)}$. This estimate is taken from \cite{Berthier:2015gja}. It represents a suppression of the SM prediction by $(v/\Lambda)^4$, where the vacuum expectation value $v=246$GeV and $\Lambda\approx 1$TeV. We implement this simply by adding $\left(10^{-3} O^{(0)}_i \right)^2$ to the diagonal element $V_{ii}$ of the covariance matrix in the following function, which we then treat as a $\Delta\chi^2$,
\bea
\left( \hat O - O^{(0)} - O^{(1)} C \right)^T V^{-1}  \left( \hat O - O^{(0)} - O^{(1)} C \right)	\,.
\eea

Figure~\ref{n_Dependency} shows rather loose constraints on $C_W$. With our proposed method the $\Delta\chi^2$ changes rather dramatically, and the $3$ and $4\sigma$ confidence intervals widen significantly. In addition we see here the effect of poor estimates of the theoretical error. Multiplying our estimate of the uncertainty by 2, the confidence interval changes significantly at $3\sigma$, and catastrophically at $4\sigma$. Putting a permille level constant error changes almost nothing for $C_W$. Figure~\ref{SMEFT_Dependency} on the other hand shows a very tightly constrained parameter $C_{lq}^{(3)}$. With such a small parameter, the inclusion of second order uncertainties changes almost nothing. The permille constant error is now relatively large, and the changes to the confidence intervals are enormous. Figure~\ref{Comparison_Method} finally shows the coefficient $C_{lu}$, for which both methods show visible but small changes. 

The overall effect of our method as seen here is easily understood. It exactly accounts for the fact that small parameters lead to small uncertainties, and large parameters give large uncertainties. We are not in charge of inserting by hand the appropriate Wilson coefficients, only the factors in front. The opposite effect is seen when putting a constant error which implies we know the correct Wilson coefficient beforehand. This exactly leads to the small parameters having overestimated errors and vice versa. 

\begin{figure}[t]
\centering
\includegraphics[width=0.82 \textwidth]{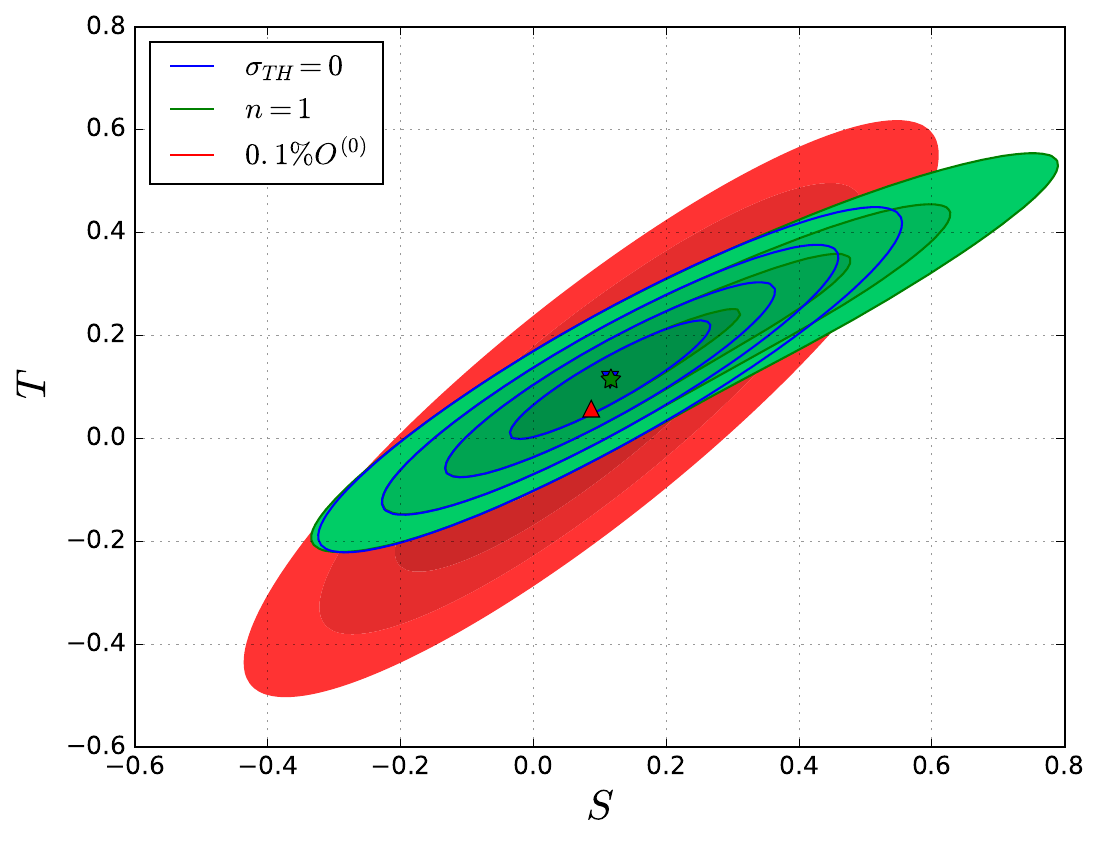}
\caption{$1, 2, 3$ and $4\sigma$ confidence regions of the $S$ and $T$ parameters, defined in Eq.~\eqref{eq:STdef}, in blue line when no theoretical error is added, in green shades when we add a theoretical uncertainty $ \sigma_{TH, ijk}$ and in red shades when a constant theoretical error of $0.1\%$ of the SM prediction is added. The minima of the functions are shown in stars and triangles in the same colors. All other Wilson coefficients are taken to be 0. \label{ST}}
\end{figure}

The difference between the two methods becomes more clear in a 2D plot. We take here the oblique parameters $S$ and $T$, defined as 
\bea\label{eq:STdef}
S = \frac{16 \pi v^2}{g_1 g_2}\frac{C_{HWB}}{\Lambda^2}, \hspace{0.5cm} T = -2\pi v^2\left(\frac{g_1^2 + g_2^2}{g_1^2 g_2^2}\right)\frac{C_{HD}}{\Lambda^2}
\eea
where $g_1=0.35$ and $g_2=0.65$ are the gauge couplings of $U(1)_Y$ and $SU(2)_L$. Figure~\ref{ST} shows the constraints on these two parameters with all other Wilson coefficients set to zero. Using a constant error misestimates not just the size of the contours but also the direction of the parameter degeneracy. An effect of this is a changing minimum for eg. $S$ when keeping $T=0$ fixed like the rest of the parameters. With no extra error the minimum of the function is clearly negative, while after adding a constant error, it shifts to a positive value. This effect can also be seen for $C_W$ in Figure~\ref{SMEFT_Dependency}.

The major difference with our method is that as the coefficients grow, they become harder and harder to constrain since the errors grow as well. This is why the contours are dragged towards the upper-right corner of high $S$ and $T$. Notice however, that around the origin the blue and green contours coincide. This effect is not seen in the lower left corner since the values of $|S|$ and $|T|$ there are too small to change the $3$- and $4\sigma$ contours noticeably. 

\section{Conclusion}\label{sec:con}
In this work, we present a new method to fit for small parameters when theoretical predictions are missing higher order corrections in the parameters. Our method relies on the assumption that the coefficients of the higher order corrections can be treated as gaussian distributed random variables. It takes advantage of the fact that in a frequentist setting, the only relevant quantity is the distribution of the statistic evaluated at the true parameters. This allows us to construct the exact $\Delta\chi^2$ statistic at the true parameters, even without knowing beforehand what they are. The proposed $\Delta\chi^2$ statistic in Eq.~\eqref{eq:stat} is chi-squared distributed by construction, which is almost surely not the case when adding a fixed or no theoretical error.

Furthermore we apply this method to fit for the SMEFT parameters associated with dimension-6 operators. We find that it impacts mainly the confidence regions of poorly constrained parameters, while leaving the ones of very constrained parameters almost unchanged.
The reason for this is that larger parameters lead to more uncertain theoretical predictions, which in turn makes the constraints weaker, and enlarges the confidence regions. We compare our results to the ones obtained when adding a constant error to account for these missing higher orders in the fitted parameters. The disagreements we find between the two methods show how including a constant theoretical error misestimates the confidence regions of the relevant parameters.  

We hope our result may open new avenues for future work. The crude estimates we do here for the SMEFT fit can be improved in a more careful analysis, taking into account both theoretical correlations and the nature of the observables. As any perturbative theory has errors of this sort, we find that it is important to estimate and treat these uncertainties correctly, especially in a time of precision physics.

\section*{Acknowledgments}
We would like to thank Alexander Karlberg, Troels Petersen and J{\o}rgen Beck Hansen for interested reading and helpful comments on the manuscript. We thank the SciPost referees for constructive comments on the manuscript.
\paragraph{Funding information}
This work was supported by Danmarks Grundforskningsfond under grant no. 1041811001.


\begin{thebibliography}{10}
\providecommand{\url}[1]{\texttt{#1}}
\providecommand{\urlprefix}{URL }
\expandafter\ifx\csname urlstyle\endcsname\relax
  \providecommand{\doi}[1]{doi:\discretionary{}{}{}#1}\else
  \providecommand{\doi}{doi:\discretionary{}{}{}\begingroup
  \urlstyle{rm}\Url}\fi
\providecommand{\eprint}[2][]{\url{#2}}

\bibitem{Aad:2012tfa}
G.~Aad \emph{et~al.},
\newblock \emph{{Observation of a new particle in the search for the Standard
  Model Higgs boson with the ATLAS detector at the LHC}},
\newblock Phys. Lett. \textbf{B716}, 1 (2012),
\newblock \doi{10.1016/j.physletb.2012.08.020},
\newblock \eprint{1207.7214}.

\bibitem{Chatrchyan:2012xdj}
S.~Chatrchyan \emph{et~al.},
\newblock \emph{{Observation of a new boson at a mass of 125 GeV with the CMS
  experiment at the LHC}},
\newblock Phys. Lett. \textbf{B716}, 30 (2012),
\newblock \doi{10.1016/j.physletb.2012.08.021},
\newblock \eprint{1207.7235}.

\bibitem{Peskin:1990zt}
M.~E. Peskin and T.~Takeuchi,
\newblock \emph{{A New constraint on a strongly interacting Higgs sector}},
\newblock Phys. Rev. Lett. \textbf{65}, 964 (1990),
\newblock \doi{10.1103/PhysRevLett.65.964}.

\bibitem{Peskin:1991sw}
M.~E. Peskin and T.~Takeuchi,
\newblock \emph{{Estimation of oblique electroweak corrections}},
\newblock Phys.Rev. \textbf{D46}, 381 (1992),
\newblock \doi{10.1103/PhysRevD.46.381}.

\bibitem{Kennedy:1991sn}
D.~C. Kennedy and P.~Langacker,
\newblock \emph{{Precision electroweak experiments and heavy physics: An
  Update}},
\newblock Phys. Rev. \textbf{D44}, 1591 (1991),
\newblock \doi{10.1103/PhysRevD.44.1591}.

\bibitem{Aad:2015pla}
G.~Aad \emph{et~al.},
\newblock \emph{{Constraints on new phenomena via Higgs boson couplings and
  invisible decays with the ATLAS detector}},
\newblock JHEP \textbf{11}, 206 (2015),
\newblock \doi{10.1007/JHEP11(2015)206},
\newblock \eprint{1509.00672}.

\bibitem{Aad:2015tna}
G.~Aad \emph{et~al.},
\newblock \emph{{Constraints on non-Standard Model Higgs boson interactions in
  an effective Lagrangian using differential cross sections measured in the $H
  \rightarrow \gamma\gamma$ decay channel at $\sqrt{s} = 8$TeV with the ATLAS
  detector}},
\newblock Phys. Lett. \textbf{B753}, 69 (2016),
\newblock \doi{10.1016/j.physletb.2015.11.071},
\newblock \eprint{1508.02507}.

\bibitem{Khachatryan:2016tnr}
V.~Khachatryan \emph{et~al.},
\newblock \emph{{Combined search for anomalous pseudoscalar HVV couplings in
  VH(H $\to b \bar b$) production and H $\to$ VV decay}},
\newblock Phys. Lett. \textbf{B759}, 672 (2016),
\newblock \doi{10.1016/j.physletb.2016.06.004},
\newblock \eprint{1602.04305}.

\bibitem{Aad:2016sau}
G.~Aad \emph{et~al.},
\newblock \emph{{Measurements of $Z\gamma$ and $Z\gamma\gamma$ production in
  $pp$ collisions at $\sqrt{s}=$ 8 TeV with the ATLAS detector}},
\newblock Phys. Rev. \textbf{D93}(11), 112002 (2016),
\newblock \doi{10.1103/PhysRevD.93.112002},
\newblock \eprint{1604.05232}.

\bibitem{Aad:2016ett}
G.~Aad \emph{et~al.},
\newblock \emph{{Measurements of $W^\pm Z$ production cross sections in $pp$
  collisions at $\sqrt{s} = 8$ TeV with the ATLAS detector and limits on
  anomalous gauge boson self-couplings}},
\newblock Phys. Rev. \textbf{D93}(9), 092004 (2016),
\newblock \doi{10.1103/PhysRevD.93.092004},
\newblock \eprint{1603.02151}.

\bibitem{Aad:2016wpd}
G.~Aad \emph{et~al.},
\newblock \emph{{Measurement of total and differential $W^+W^-$ production
  cross sections in proton-proton collisions at $\sqrt{s}=$ 8 TeV with the
  ATLAS detector and limits on anomalous triple-gauge-boson couplings}}
  (2016),
\newblock \eprint{1603.01702}.

\bibitem{Khachatryan:2015kea}
V.~Khachatryan \emph{et~al.},
\newblock \emph{{Measurement of the Z$\gamma$ Production Cross Section in pp
  Collisions at 8 TeV and Search for Anomalous Triple Gauge Boson Couplings}},
\newblock JHEP \textbf{04}, 164 (2015),
\newblock \doi{10.1007/JHEP04(2015)164},
\newblock \eprint{1502.05664}.

\bibitem{Khachatryan:2015sga}
V.~Khachatryan \emph{et~al.},
\newblock \emph{{Measurement of the ${{\mathrm{W} }^{+} }\mathrm{W}^{-} $ cross
  section in pp collisions at $\sqrt{s} = 8$ TeV and limits on anomalous gauge
  couplings}},
\newblock Eur. Phys. J. \textbf{C76}(7), 401 (2016),
\newblock \doi{10.1140/epjc/s10052-016-4219-1},
\newblock \eprint{1507.03268}.

\bibitem{Khachatryan:2016yro}
V.~Khachatryan \emph{et~al.},
\newblock \emph{{Measurement of the $ \mathrm{ Z } \gamma \rightarrow \nu
  \bar{\nu} \gamma$ production cross section in pp collisions at $\sqrt{s}=$ 8
  TeV and limits on anomalous $ \mathrm{ ZZ } \gamma$ and $ \mathrm{Z} \gamma
  \gamma$ trilinear gauge boson couplings}},
\newblock Phys. Lett. \textbf{B760}, 448 (2016),
\newblock \doi{10.1016/j.physletb.2016.06.080},
\newblock \eprint{1602.07152}.

\bibitem{Grinstein:1991cd}
B.~Grinstein and M.~B. Wise,
\newblock \emph{{Operator analysis for precision electroweak physics}},
\newblock Phys. Lett. \textbf{B265}, 326 (1991),
\newblock \doi{10.1016/0370-2693(91)90061-T}.

\bibitem{Han:2004az}
Z.~Han and W.~Skiba,
\newblock \emph{{Effective theory analysis of precision electroweak data}},
\newblock Phys. Rev. \textbf{D71}, 075009 (2005),
\newblock \doi{10.1103/PhysRevD.71.075009},
\newblock \eprint{hep-ph/0412166}.

\bibitem{Berthier:2015gja}
L.~Berthier and M.~Trott,
\newblock \emph{{Consistent constraints on the Standard Model Effective Field
  Theory}},
\newblock JHEP \textbf{02}, 069 (2016),
\newblock \doi{10.1007/JHEP02(2016)069},
\newblock \eprint{1508.05060}.

\bibitem{Berthier:2015oma}
L.~Berthier and M.~Trott,
\newblock \emph{{Towards consistent Electroweak Precision Data constraints in
  the SMEFT}},
\newblock JHEP \textbf{05}, 024 (2015),
\newblock \doi{10.1007/JHEP05(2015)024},
\newblock \eprint{1502.02570}.

\bibitem{Berthier:2016tkq}
L.~Berthier, M.~Bj{\o}rn and M.~Trott,
\newblock \emph{{Incorporating doubly resonant $W^\pm$ data in a global fit of
  SMEFT parameters to lift flat directions}}  (2016),
\newblock \eprint{1606.06693}.

\bibitem{Dumont:2013wma}
B.~Dumont, S.~Fichet and G.~von Gersdorff,
\newblock \emph{{A Bayesian view of the Higgs sector with higher dimensional
  operators}},
\newblock JHEP \textbf{07}, 065 (2013),
\newblock \doi{10.1007/JHEP07(2013)065},
\newblock \eprint{1304.3369}.

\bibitem{Ciuchini:2014dea}
M.~Ciuchini, E.~Franco, S.~Mishima, M.~Pierini, L.~Reina and L.~Silvestrini,
\newblock \emph{{Update of the electroweak precision fit, interplay with
  Higgs-boson signal strengths and model-independent constraints on new
  physics}},
\newblock In \emph{{Proceedings, 37th International Conference on High Energy
  Physics (ICHEP 2014): Valencia, Spain, July 2-9, 2014}},
\newblock \doi{10.1016/j.nuclphysbps.2015.09.361} (2016), \eprint{1410.6940}.

\bibitem{Falkowski:2015fla}
A.~Falkowski,
\newblock \emph{{Effective field theory approach to LHC Higgs data}}  (2015),
\newblock \eprint{1505.00046}.

\bibitem{Falkowski:2015jaa}
A.~Falkowski, M.~Gonzalez-Alonso, A.~Greljo and D.~Marzocca,
\newblock \emph{{Global constraints on anomalous triple gauge couplings in
  effective field theory approach}},
\newblock Phys. Rev. Lett. \textbf{116}(1), 011801 (2016),
\newblock \doi{10.1103/PhysRevLett.116.011801},
\newblock \eprint{1508.00581}.

\bibitem{Buckley:2015nca}
A.~Buckley, C.~Englert, J.~Ferrando, D.~J. Miller, L.~Moore, M.~Russell and
  C.~D. White,
\newblock \emph{{Global fit of top quark effective theory to data}},
\newblock Phys. Rev. \textbf{D92}(9), 091501 (2015),
\newblock \doi{10.1103/PhysRevD.92.091501},
\newblock \eprint{1506.08845}.

\bibitem{Buckley:2015lku}
A.~Buckley, C.~Englert, J.~Ferrando, D.~J. Miller, L.~Moore, M.~Russell and
  C.~D. White,
\newblock \emph{{Constraining top quark effective theory in the LHC Run II
  era}},
\newblock JHEP \textbf{04}, 015 (2016),
\newblock \doi{10.1007/JHEP04(2016)015},
\newblock \eprint{1512.03360}.

\bibitem{Gregersen:2015uea}
K.~D. Gregersen and J.~B. Hansen,
\newblock \emph{{Frequentist limit setting in effective field theories}}
  (2015),
\newblock \eprint{1509.01808}.

\bibitem{Bagnaschi:2014wea}
E.~Bagnaschi, M.~Cacciari, A.~Guffanti and L.~Jenniches,
\newblock \emph{{An extensive survey of the estimation of uncertainties from
  missing higher orders in perturbative calculations}},
\newblock JHEP \textbf{02}, 133 (2015),
\newblock \doi{10.1007/JHEP02(2015)133},
\newblock \eprint{1409.5036}.

\bibitem{Cacciari:2011ze}
M.~Cacciari and N.~Houdeau,
\newblock \emph{{Meaningful characterisation of perturbative theoretical
  uncertainties}},
\newblock JHEP \textbf{09}, 039 (2011),
\newblock \doi{10.1007/JHEP09(2011)039},
\newblock \eprint{1105.5152}.

\bibitem{Isgro:2015cfa}
A.~Isgr{\`o},
\newblock \emph{{Theoretical uncertainties on QCD computations and the
  divergence of the perturbative expansion}},
\newblock Ph.D. thesis, Milan U. (2015), \eprint{1509.00359}.

\bibitem{Furnstahl:2015rha}
R.~J. Furnstahl, N.~Klco, D.~R. Phillips and S.~Wesolowski,
\newblock \emph{{Quantifying truncation errors in effective field theory}},
\newblock Phys. Rev. \textbf{C92}(2), 024005 (2015),
\newblock \doi{10.1103/PhysRevC.92.024005},
\newblock \eprint{1506.01343}.

\bibitem{Fichet:2015xla}
S.~Fichet and G.~Moreau,
\newblock \emph{{Anatomy of the Higgs fits: a first guide to statistical
  treatments of the theoretical uncertainties}},
\newblock Nucl. Phys. \textbf{B905}, 391 (2016),
\newblock \doi{10.1016/j.nuclphysb.2016.02.019},
\newblock \eprint{1509.00472}.

\bibitem{Grzadkowski:2010es}
B.~Grzadkowski, M.~Iskrzynski, M.~Misiak and J.~Rosiek,
\newblock \emph{{Dimension-Six Terms in the Standard Model Lagrangian}},
\newblock JHEP \textbf{1010}, 085 (2010),
\newblock \doi{10.1007/JHEP10(2010)085},
\newblock \eprint{1008.4884}.

\bibitem{ALEPH:2005ab}
S.~Schael \emph{et~al.},
\newblock \emph{{Precision electroweak measurements on the $Z$ resonance}},
\newblock Phys. Rept. \textbf{427}, 257 (2006),
\newblock \doi{10.1016/j.physrep.2005.12.006},
\newblock \eprint{hep-ex/0509008}.

\bibitem{Schael:2013ita}
S.~Schael \emph{et~al.},
\newblock \emph{{Electroweak Measurements in Electron-Positron Collisions at
  W-Boson-Pair Energies at LEP}},
\newblock Phys. Rept. \textbf{532}, 119 (2013),
\newblock \doi{10.1016/j.physrep.2013.07.004},
\newblock \eprint{1302.3415}.

\bibitem{Inoue:2000hc}
Y.~Inoue \emph{et~al.},
\newblock \emph{{Measurement of the cross-section and forward - backward charge
  asymmetry for the b and c quark in e+ e- annihilation with inclusive muons at
  s**(1/2) = 58-GeV}},
\newblock Eur. Phys. J. \textbf{C18}, 273 (2000),
\newblock \doi{10.1007/s100520000541},
\newblock \eprint{hep-ex/0012033}.

\bibitem{Velissaris:1994rv}
C.~Velissaris \emph{et~al.},
\newblock \emph{{Measurements of cross-section and charge asymmetry for e+ e-
  ---> mu+ mu- and e+ e- ---> tau+ tau- at s**(1/2) = 57.8-GeV}},
\newblock Phys.Lett. \textbf{B331}, 227 (1994),
\newblock \doi{10.1016/0370-2693(94)90967-9}.

\bibitem{Miura:1997mq}
M.~Miura \emph{et~al.},
\newblock \emph{{Precise measurement of the e+ e- ---> mu+ mu- reaction at
  s**(1/2) = 57.77-GeV}},
\newblock Phys.Rev. \textbf{D57}, 5345 (1998),
\newblock \doi{10.1103/PhysRevD.57.5345}.

\bibitem{Hegner:1989rd}
S.~Hegner \emph{et~al.},
\newblock \emph{{Final Results on $\mu$ and Tau Pair Production by the Jade
  Collaboration at {PETRA}}},
\newblock Z.Phys. \textbf{C46}, 547 (1990),
\newblock \doi{10.1007/BF01560255}.

\bibitem{Derrick:1985gs}
M.~Derrick, E.~Fernandez, R.~Fries, L.~Hyman, P.~Kooijman \emph{et~al.},
\newblock \emph{{New Results on the Reaction $e^+ e^- \to \mu^+ \mu^-$ at
  $\sqrt{s}=29$-{GeV}}},
\newblock Phys.Rev. \textbf{D31}, 2352 (1985),
\newblock \doi{10.1103/PhysRevD.31.2352}.

\bibitem{Sagawa:1996yf}
H.~Sagawa, T.~Tauchi, M.~Tanabashi and S.~Uehara,
\newblock \emph{{TRISTAN physics at high luminosities. Proceedings, 3rd
  Workshop, Tsukuba, Japan, November 16-18, 1994}}  (1996).

\bibitem{Abe:1993xr}
K.~Abe \emph{et~al.},
\newblock \emph{{A Study of the charm and bottom quark production in e+ e-
  annihilation at s**(1/2) = 58-GeV using prompt electrons}},
\newblock Phys.Lett. \textbf{B313}, 288 (1993),
\newblock \doi{10.1016/0370-2693(93)91226-D}.

\bibitem{Gan:1985st}
K.~K. Gan \emph{et~al.},
\newblock \emph{{Measurement of the Reaction $e^+ e^- \to \tau^+ \tau^-$ at
  $\sqrt{s}=29$-{GeV}}},
\newblock Phys. Lett. \textbf{B153}, 116 (1985),
\newblock \doi{10.1016/0370-2693(85)91453-4}.

\bibitem{Zeller:2001hh}
G.~Zeller \emph{et~al.},
\newblock \emph{{A Precise determination of electroweak parameters in neutrino
  nucleon scattering}},
\newblock Phys.Rev.Lett. \textbf{88}, 091802 (2002),
\newblock \doi{10.1103/PhysRevLett.88.091802},
\newblock \eprint{hep-ex/0110059}.

\bibitem{Allaby:1987vr}
J.~Allaby \emph{et~al.},
\newblock \emph{{A Precise Determination of the Electroweak Mixing Angle from
  Semileptonic Neutrino Scattering}},
\newblock Z.Phys. \textbf{C36}, 611 (1987),
\newblock \doi{10.1007/BF01630598}.

\bibitem{Vilain:1994qy}
P.~Vilain \emph{et~al.},
\newblock \emph{{Precision measurement of electroweak parameters from the
  scattering of muon-neutrinos on electrons}},
\newblock Phys.Lett. \textbf{B335}, 246 (1994),
\newblock \doi{10.1016/0370-2693(94)91421-4}.

\bibitem{Ahrens:1990fp}
L.~A. Ahrens \emph{et~al.},
\newblock \emph{{Determination of electroweak parameters from the elastic
  scattering of muon-neutrinos and anti-neutrinos on electrons}},
\newblock Phys. Rev. \textbf{D41}, 3297 (1990),
\newblock \doi{10.1103/PhysRevD.41.3297}.

\bibitem{McFarland:1997wx}
K.~S. McFarland \emph{et~al.},
\newblock \emph{{A Precision measurement of electroweak parameters in neutrino
  - nucleon scattering}},
\newblock Eur.Phys.J. \textbf{C1}, 509 (1998),
\newblock \doi{10.1007/s100520050099},
\newblock \eprint{hep-ex/9701010}.

\bibitem{Anthony:2005pm}
P.~Anthony \emph{et~al.},
\newblock \emph{{Precision measurement of the weak mixing angle in Moller
  scattering}},
\newblock Phys.Rev.Lett. \textbf{95}, 081601 (2005),
\newblock \doi{10.1103/PhysRevLett.95.081601},
\newblock \eprint{hep-ex/0504049}.

\bibitem{Prescott:1979dh}
C.~Prescott, W.~Atwood, R.~L. Cottrell, H.~DeStaebler, E.~L. Garwin
  \emph{et~al.},
\newblock \emph{{Further Measurements of Parity Nonconservation in Inelastic
  electron Scattering}},
\newblock Phys.Lett. \textbf{B84}, 524 (1979),
\newblock \doi{10.1016/0370-2693(79)91253-X}.

\bibitem{Dorenbosch:1988is}
J.~Dorenbosch \emph{et~al.},
\newblock \emph{{EXPERIMENTAL RESULTS ON NEUTRINO - ELECTRON SCATTERING}},
\newblock Z. Phys. \textbf{C41}, 567 (1989),
\newblock \doi{10.1007/BF01564701},
\newblock [Erratum: Z. Phys.C51,142(1991)].

\bibitem{Vetter:1995vf}
P.~Vetter, D.~Meekhof, P.~Majumder, S.~Lamoreaux and E.~Fortson,
\newblock \emph{{Precise test of electroweak theory from a new measurement of
  parity nonconservation in atomic thallium}},
\newblock Phys.Rev.Lett. \textbf{74}, 2658 (1995),
\newblock \doi{10.1103/PhysRevLett.74.2658}.

\bibitem{Wood:1997zq}
C.~Wood, S.~Bennett, D.~Cho, B.~Masterson, J.~Roberts \emph{et~al.},
\newblock \emph{{Measurement of parity nonconservation and an anapole moment in
  cesium}},
\newblock Science \textbf{275}, 1759 (1997),
\newblock \doi{10.1126/science.275.5307.1759}.

\bibitem{Ito:2003mr}
T.~Ito \emph{et~al.},
\newblock \emph{{Parity violating electron deuteron scattering and the proton's
  neutral weak axial vector form-factor}},
\newblock Phys.Rev.Lett. \textbf{92}, 102003 (2004),
\newblock \doi{10.1103/PhysRevLett.92.102003},
\newblock \eprint{nucl-ex/0310001}.

\bibitem{Achard:2004zw}
P.~Achard \emph{et~al.},
\newblock \emph{{Measurement of the cross section of W-boson pair production at
  LEP}},
\newblock Phys. Lett. \textbf{B600}, 22 (2004),
\newblock \doi{10.1016/j.physletb.2004.08.060},
\newblock \eprint{hep-ex/0409016}.

\bibitem{Abbiendi:2007rs}
G.~Abbiendi \emph{et~al.},
\newblock \emph{{Measurement of the e+ e- ---> W+ W- cross section and W decay
  branching fractions at LEP}},
\newblock Eur. Phys. J. \textbf{C52}, 767 (2007),
\newblock \doi{10.1140/epjc/s10052-007-0442-0},
\newblock \eprint{0708.1311}.

\bibitem{Heister:2004wr}
A.~Heister \emph{et~al.},
\newblock \emph{{Measurement of W-pair production in e+ e- collisions at
  centre-of-mass energies from 183-GeV to 209-GeV}},
\newblock Eur. Phys. J. \textbf{C38}, 147 (2004),
\newblock \doi{10.1140/epjc/s2004-02048-3}.

\end{thebibliography}

\nolinenumbers

\end{document}